\documentclass[12pt]{article}%
\usepackage{amsmath}
\usepackage{amsfonts}
\usepackage{amssymb}
\usepackage{graphicx}
\usepackage{hyperref}%
\begin{document}

\title{The quantization of gravity and the vacuum energy of quantum fields}
\author{M. Chaves\\\textit{Universidad de Costa Rica}\\\textit{San Jos\'{e}, Costa Rica}\\mchaves@cariari.ucr.ac.cr}
\date{June 25, 2007}
\maketitle

\begin{abstract}
We construct a unified covariant derivative that contains the sum of an affine
connection and a Yang-Mills field. With it we construct a lagrangian that is
invariant both under diffeomorphisms and Yang-Mills gauge transformations. We
assume that metric and symmetric affine connection are independent quantities,
and make the observation that the metric must be able to generate curvature,
just as the connection, so there should be an extra tensor similar to
Riemann's in the equations but constructed from metrics and not connections.
We find the equations generated by the lagrangian and introduce the huge
natural scale due to the vacuum energy of quantum fields. This scale allows
for a perturbative solution of the equations of motion. We prove the system
has a vacuum state that forces the metricity of the metric and results in
General Relativity for low energies. The vacuum energy of the quantum fields
cancels, becoming unobservable. At very high energies, the metric does not
appear differentiated in the lagrangian and so it is not a quantum field, just
a background classical field. The true quantum fields are the connections. The
theory becomes very similar to a Yang-Mills, with affine connections taking
the place of Yang-Mills vector fields. It should be renormalizable since it
has a coupling constant with no units and correct propagators after fixing the
gauge (diffeomorphisms). The weakness of gravity turns out to be due to the
large vacuum energy of the quantum fields.

\end{abstract}

\section{INTRODUCTION.}

The force of gravity has been very hard to understand at the quantum level.
The attempts at a covariant quantization of \b{G}eneral \b{R}elativity (GR)
present the difficulty that the coupling constant is $\kappa=\sqrt{32\pi
G}\sim1/E_{P},$ where $E_{P}$ is Planck's energy and $G$ is the gravitational
constant. A coupling constant with units leads to an unrenormalizable quantum
field theory since it requires an infinite number of different counterterms
for the quantum loops. This is also true in supergravity.\cite{Deser} This
limits somewhat the theory's usefulness although it is still possible to
obtain interesting\cite{correction} results seeing it as an effective field
theory.\cite{EFT} At any rate this situation is a clear reminder of how
limited our understanding of gravitation is.

In GR the gravitational term in the lagrangian density is proportional to the
scalar of curvature, that is, $\mathcal{L}_{G}\varpropto G^{-1}R.$ Thus the
gravitational term has the units $E^{4},$ which are the correct ones for four
dimensions$.$ In the case of Yang-Mills Theories the kinetic energy in the
lagrangian has the form $F^{2},$ where $F$ is the two-form curvature
associated with the Yang-Mills connection $A.$ As a matter of fact $F_{\mu\nu
}=[D_{\mu},D_{\nu}],$ where $D_{\mu}=\partial_{\mu}+A_{\mu}.$ The units of
$F^{2}$ are $E^{4},$ as in the case of the gravitational term, but now the
coupling constant (hidden in the definition of $A$ and given below) has not
units. Thus we see that the cause of our troubles with gravitation is that the
curvature appears to the first power in the lagrangian, and we must substitute
the other curvature factor with a constant with units. It would be ideal if
the gravitational term would emulate Yang-Mills Theories and have a form
quadratic in the curvature. There have been many efforts to construct a gauge
theory of gravity, often gauging the Poincar\'{e} group. The most common cause
these efforts have not been satisfying are the difficulties involved in
obtaining GR, that has to appear as a limit of some kind if we have any
expectations for our theory to be realistic.

The guiding idea of this paper is that, although we really know very little
about the nature of either GR or Yang-Mills Theories, we do have a clear idea
of how to algebraically manipulate these two theories and what they are about:
gauge invariance. Our aim is to take an intermediate step where we still
maintain covariance (diffeomorphic invariance) and Yang-Mills invariance
(local Lie group invariance) by means of an unified covariant derivative. This
derivative is simply the sum of the Yang-Mills and the affine connections and
by itself contains no new physics. But with it we can construct a curvature,
and use the square of this curvature as lagrangian. The coupling constant of a
theory of this kind has no units, and quantization of the theory should
present no problems. But how to obtain now GR?

We assume the independence of a \emph{symmetric} (following present
understanding\cite{Querella}) connection and the metric, in order to obtain a
$A_{n}$ manifold. Thus there is zero torsion and no metricity of the metric.
The metricity of the metric, or \emph{metricity},\emph{ }for short, is the
condition%
\begin{equation}
g_{\mu\nu;\lambda}=0. \label{metricity}%
\end{equation}
We use this independence when we obtain the equations of motion from the
lagrangian, but also in a more direct and fundamental way. Thus we make the
observation that if the connection and the metric are truly independent
objects, then each one should be able, in principle, to produce its own
distinct contribution to the curvature of spacetime. The curvature due to the
affine connection can be expressed by the familiar Riemann tensor:%
\begin{equation}
R^{\rho}{}_{\sigma\mu\nu}[\Gamma]=\Gamma_{\sigma\mu}{}^{\rho}{}_{,\nu}%
-\Gamma_{\sigma\nu}{}^{\rho}{}_{,\mu}+\Gamma_{\sigma\mu}{}^{\tau}\Gamma
_{\nu\tau}{}^{\rho}-\Gamma_{\sigma\nu}{}^{\tau}\Gamma_{\mu\tau}{}^{\rho},
\label{Riemann1}%
\end{equation}
which is a functional of the connection (but not of the metric). The curvature
due to the metric is expressed by what we shall call the \b{m}etric \b
{c}urvature \b{t}ensor (MCT):%
\begin{equation}
\bar{R}^{\rho}{}_{\sigma\mu\nu}[g]=K(\delta_{\nu}{}^{\rho}g_{\sigma\mu}%
-\delta_{\mu}{}^{\rho}g_{\sigma\nu}), \label{Riemann2}%
\end{equation}
which is a functional of the metric (but not of the connection). We then
postulate that the total curvature $\hat{R}^{\rho}{}_{\sigma\mu\nu}$ of the
manifold is given by%
\begin{equation}
\hat{R}^{\rho}{}_{\sigma\mu\nu}=R^{\rho}{}_{\sigma\mu\nu}+\bar{R}^{\rho}%
{}_{\sigma\mu\nu}. \label{Riemann12}%
\end{equation}
The introduction of this second curvature tensor has very interesting and
unexpected consequences, that arise from its special algebraic properties.

Our observational spacetime is a degenerate type of manifold compared to the
one we are postulating, since in it the metric and the connection are
dependent. But it is still possible to use either the connection or the metric
to measure curvature in it. We can use a connection to study how a vector
rotates when it is displaced parallel to itself around a closed path lying on
the surface. Alternatively, we can also measure the surface's curvature using
the metric, as follows: we take concentric circles and measure their
radius-to-circumference ratio. If for a circle $r/C=1/2\pi$, then the surface
is flat; if $2\pi r/C>1/2\pi$, then it is hyperbolic, and if $2\pi r/C<1/2\pi
$, elliptical.

The existence of a curvature due to the metric is really the only hypothesis
of the paper with new content. The other one we shall require, that quantum
fields have a very large vacuum energy density $\rho_{0}$, is basically
accepted, the peculiar thing being that this density is not observed. If it
were, the universe would be the size of a pea. We obtain the equations of
motion generated by the lagrangian, and some interesting things happen: the
large scale given by the vacuum energy of the quantum fields allows a
perturbative solution of the equations of motion. This solution has a vacuum
that makes the theory resemble GR. The small value of $G,$ the gravitational
constant, is due directly to the large value of the energy density scale, as
it turns out that $G\sim\rho_{0}^{-1/2}.$ Also, very conveniently, the vacuum
energy density $\rho_{0}$ cancels out whatever its original may be.

As long as the energy is small, that is, as long as $E\lesssim E_{P}$ holds,
the theory resembles GR. However, if the energies involved grow larger so that
$E\gg E_{P}$ holds, it is possible to neglect the MCT term and we are back to
a Yang-Mills-like theory, with a coupling constant with no units and a
promising Feynman diagram structure. The theory should probably be renormalizable.

In Section 2 we establish the mathematical conventions, definitions and
identities we shall need; in Section 3 we define a unified covariant
derivative that has both Yang-Mills and affine connections, and with it
construct the lagrangian, adding, too, the MCT term; in Section 4 we solve the
equations of motion and find the vacuum state of the theory; in Section 5 we
study the renormalization of gravity for very high energies; in Section 6 we
present our final comments. At the end there is an Appendix, where the
derivation of the equations of motion from the lagrangian is done in detail.

\section{Mathematical conventions, definitions and identities.}

The conventions, definitions and identities are related to GR, Yang-Mills
Theories and the MCT.

\subsection{The conventions, definitions and identities related to General
Relativity (GR).}

Most of the calculations will be done on an $n$-dimensional spacetime. The
spacetime indices are to be represented by Greek letters later in the
alphabet: $\lambda,\mu,\nu,\xi,\ldots$, with the metric $g_{\mu\nu},$ $\mu
,\nu=0,1,2,\ldots,n-1$ having the signature $(-+++\cdots).$ We use
$e\equiv\sqrt{-\det(g_{\mu\nu})}$ and define the Riemann tensor as in
(\ref{Riemann1}):%
\begin{align*}
R^{\rho}{}_{\sigma\mu\nu}  &  =\Gamma_{\sigma\mu}{}^{\rho}{}_{,\nu}%
-\Gamma_{\sigma\nu}{}^{\rho}{}_{,\mu}+\Gamma_{\sigma\mu}{}^{\tau}\Gamma
_{\nu\tau}{}^{\rho}-\Gamma_{\sigma\nu}{}^{\tau}\Gamma_{\mu\tau}{}^{\rho}\\
&  =\Gamma_{\sigma\lbrack\mu}{}^{\rho}{}_{,\nu]}+\Gamma_{\sigma\lbrack\mu}%
{}^{\tau}\Gamma_{\nu]\tau}{}^{\rho}.
\end{align*}
The Ricci tensor is $R_{\sigma\nu}=R^{\rho}{}_{\sigma\rho\nu}$ and its
contraction is the curvature scalar $R.$ As usual we take the stress-energy
tensor to be%
\begin{equation}
T_{\mu\nu}=-\frac{2}{e}\frac{\delta I_{M}}{\delta g^{\mu\nu}}, \label{stress}%
\end{equation}
where $I_{M}$ is the matter lagrangian$.$ We note that $\delta e=-\frac{1}%
{2}eg_{\mu\nu}\delta g^{\mu\nu}$.

The indices of the coordinates of a tangential flat space at a point of the
base spacetime manifold will be taken from the first letters of the Greek
alphabet: $\alpha,\beta,\gamma,\delta,...$, the so-called non-holonomic
coordinates. The tetrads will thus be written $e^{\alpha}=e^{\alpha}{}_{\mu
}dx^{\mu}$ and defined by $g_{\mu\nu}=e^{\alpha}{}_{\mu}e^{\beta}{}_{\nu}%
\eta_{\alpha\beta},$ where $\eta_{\alpha\beta}=\operatorname{diag}%
(-1,1,1,1,\ldots)$ is the Minkowski metric. Thus $e=\det\nolimits^{1/2}%
(-e^{\alpha}{}_{\mu}e^{\beta}{}_{\nu}\eta_{\alpha\beta})=\det e^{\alpha}%
{}_{\mu}.$ The symbol $e_{\alpha}{}^{\mu}$ is defined by $e_{\alpha}{}^{\mu
}\equiv g^{\mu\nu}\eta_{\alpha\beta}e^{\beta}{}_{\nu}$, and has the property
that $e^{\alpha}{}_{\mu}e_{\alpha}{}^{\nu}=\delta_{\mu}{}^{\nu},$ that is, its
the inverse of the tetrad.

The Dirac matrices are defined by the algebraic condition $\gamma^{\alpha
}\gamma^{\beta}+\gamma^{\beta}\gamma^{\alpha}=2\eta^{\alpha\beta}$ and are
coordinate-independent. The matrices defined by $\gamma^{\lambda}\equiv
e_{\alpha}{}^{\lambda}\gamma^{\alpha}$ satisfy $\gamma^{\mu}\gamma^{\nu
}+\gamma^{\nu}\gamma^{\mu}=2g^{\mu\nu}.$ In terms of the tetrad,
(\ref{stress}) can be rewritten as:\cite{BD}%
\begin{equation}
T_{\mu\nu}=-\frac{e_{\alpha\mu}}{e}\frac{\delta I_{M}}{\delta e_{\alpha}%
{}^{\nu}}. \label{stress-f}%
\end{equation}

Vectors $v_{\beta}$ undergo Lorentz rotations $L_{\alpha}{}^{\beta}%
\approx\delta_{\alpha}{}^{\beta}+\omega_{\alpha}{}^{\beta}+\cdots$ in the
tangential space, where the $\omega_{\alpha}{}^{\beta}$ are the antisymmetric
rotation parameters, thus: $v_{\alpha}\rightarrow L_{\alpha}{}^{\beta}%
v_{\beta}.$ Spinors $\psi$ undergo spinor rotations $S=\exp(\frac{1}{4}%
\omega_{\alpha}{}^{\beta}\sigma^{\alpha}{}_{\beta}),$ where $\sigma^{\alpha}%
{}_{\beta}\equiv\frac{1}{2}\left(  \gamma^{\alpha}\gamma_{\beta}-\gamma
_{\beta}\gamma^{\alpha}\right)  $; thus $\psi\rightarrow S\psi.$ As an
example, if we take $\alpha,\beta=1,2,$ then $S=\exp(\frac{1}{2}\omega_{1}%
{}^{2}\sigma^{1}{}_{2})$ is the rotation due to the parameter $\omega_{1}%
{}^{2}$ about the $z$-axis.

Since these rotations depend on the coordinates $x^{\mu}$ of the point of
tangency, in order to maintain invariance of the lagragian under them we
introduce a spin connection defined in terms of the affine connection,%
\begin{equation}
\omega_{\alpha}{}^{\beta}{}_{|\mu}\equiv e_{\alpha}{}^{\lambda}e^{\beta}%
{}_{\lambda;\mu}{}=e_{\alpha}{}^{\lambda}e^{\beta}{}_{\lambda,\mu}-e_{\alpha
}{}^{\lambda}\Gamma_{\lambda\mu}{}^{\nu}e^{\beta}{}_{\nu} \label{con}%
\end{equation}
where the vertical bar emphasizes that the last subindex, unlike the next two,
is a holonomic coordinate. With the help of the spin connection, we can write
properly invariant derivatives of vectors with non-holonomic indices and
spinors. Thus for a vector field $B_{\alpha}(x)$ in the tangent space to the
manifold the covariant derivative is%
\[
D_{\mu}^{V}{}B_{\alpha}=\partial_{\mu}B_{\alpha}+\omega_{\alpha}{}^{\beta}%
{}_{|\mu}B_{\beta}.
\]
And for a spinor field $\psi(x)$ transforming in the spin representation of
the Lorentz transformation we have the covariant derivative ($F$ for fermion)%
\begin{equation}
D_{\mu}^{F}\psi\rightarrow\partial_{\mu}\psi+\Gamma_{\mu}\psi,\quad\Gamma
_{\mu}=\frac{1}{4}\sigma^{\alpha}{}_{\beta}\omega_{\alpha}{}^{\beta}{}_{|\mu}.
\label{spin con}%
\end{equation}

\subsection{The conventions, definitions and identities related to Yang-Mills
Theories.}

The unitary gauge transformations of the Yang-Mills Theory will be given by
$U(x)=e^{\Theta},$ where $\Theta=-i\Theta^{a}T^{a},$ the $T^{a}$ are the
group's generators and the $\Theta^{a}(x)$ are the group parameters. Roman
letters $a,$ $b,$ $c,$... will be used for the Yang-Mills indices. It will
often be the case that a Dirac field transforms as%
\[
\psi\rightarrow U\psi
\]
under a transformation $U$. In order to have kinetic energy terms in the
lagrangian that transform properly we need a Yang-Mills covariant derivative%
\begin{equation}
D_{\mu}^{YM}=\mathbf{1}_{N}\partial_{\mu}+A_{\mu} \label{YM derivative}%
\end{equation}
where $A_{\mu}=-igA_{\mu}^{a}(x)T^{a}$ and the $\mathbf{1}_{N}$ is an $N\times
N$ identity matrix, $N=\operatorname{Tr}\mathbf{1}_{N}$ being the
dimensionality of the fundamental representation of the gauge Lie group of the
Yang-Mills Theory. Usually this matrix is implicitly understood, but in this
paper we shall explicitly write for purposes of clarity. The $g$ is a coupling
constant with no units. We require the gauge field to transform as given by%
\[
A_{\mu}\rightarrow UA_{\mu}U^{-1}-\left(  \partial_{\mu}U\right)  U^{-1};
\]
this way the covariant derivative will transform as:%
\begin{equation}
D_{\mu}^{YM}\rightarrow UD_{\mu}^{YM}U^{-1}. \label{YM transform}%
\end{equation}
In this expression the differential operator is acting on all fields that may
be placed to its right, and not only on $U^{-1}.$ The field strength tensor
can be obtained from the covariant derivative:%
\begin{align}
G_{\mu\nu}  &  =[D_{\mu}^{YM},D_{\nu}^{YM}]=(\partial_{\mu}A_{\nu}%
)-(\partial_{\mu}A_{\nu})+[A_{\mu},A_{\nu}]\nonumber\\
&  =A_{[\nu,\mu]}+A_{[\mu}A_{\nu]}. \label{YM curvature}%
\end{align}
To work with the expression $[D_{\mu}^{YM},D_{\nu}^{YM}]$ one must assume that
they are acting on some differentiable field placed to their right, say
$[D_{\mu}^{YM},D_{\nu}^{YM}]\phi(x)$. \textit{In what follows in this paper it
shall always be assumed that such differentiable fields will always be placed
to the right of all covariant derivatives before any algebraic manipulation is
carried out.}

\subsection{The conventions, definitions and identities dealing with the
metric curvature tensor (MCT).}

The Riemann tensor carries in itself the information pertinent to the
curvature of a Riemannian manifold and is constructed using connections. It
has very specific symmetries between its indices related to its function as a
measure of curvature. If we are to construct a tensor that gives the curvature
due to the metric, there is only one way of writing it, and it is as in
(\ref{Riemann2}). We purport in this paper that in manifolds that have
independent metric and connection the total curvature of the manifold is the
sum of the curvatures due to the connection and the metric, as in
(\ref{Riemann12}).

The algebraic form of the MCT (\ref{Riemann2}) is very familiar to us. It is
the same form as that of the Riemann tensor for pseudo-Riemannian manifolds
with maximal symmetry. In such manifolds it is possible to derive this form
from general considerations regarding the Killing vectors and the maximal
symmetry. For a signature $(-+++\cdots)$ in $n$ dimensions maximally symmetric
manifolds are of basically only two types, up to diffeomorphisms, either de
Sitter ($K>0$) or anti-de Sitter ($K<0$).

Familiar maximally symmetric ($V_{n}$ Riemaniann) manifolds in have a Riemann
tensor defined in terms of the connection, which can be written in terms of
the metric. The metricity implies that a symmetric connection is Levi-Civita.
But in this paper the tensor given by (\ref{Riemann2}) is a different one,
defined in terms of metrics, not connections. It has little relation to the
Riemann tensor of a maximally symmetric manifold. It goes without saying that
the presence of the curvature term (\ref{Riemann2}) as part of the curvature
of a $A_{n}$ manifold does not imply in any way that this manifold has to be
maximally symmetric or any kind of de Sitter space.

The following are convenient definitions involving squares of the Riemann
tensor and the MCT:%
\begin{align}
\hat{S}  &  \equiv\hat{R}^{\rho}{}_{\sigma\mu\nu}g^{\mu\tau}g^{\nu\upsilon
}\hat{R}^{\sigma}{}_{\rho\tau\upsilon},\quad\hat{S}_{\mu\tau}\equiv\hat
{R}^{\rho}{}_{\sigma\mu\nu}g^{\nu\upsilon}\hat{R}^{\sigma}{}_{\rho\tau
\upsilon},\nonumber\\
S  &  \equiv R^{\rho}{}_{\sigma\mu\nu}g^{\mu\tau}g^{\nu\upsilon}R^{\sigma}%
{}_{\rho\tau\upsilon},\quad S_{\mu\tau}\equiv R^{\rho}{}_{\sigma\mu\nu}%
g^{\nu\upsilon}R^{\sigma}{}_{\rho\tau\upsilon}. \label{defs}%
\end{align}

Certain contractions involving the MCT and the Riemann tensor are going to be
useful to us, so we list them here:%
\begin{align}
\bar{R}{}_{\mu\nu}=\bar{R}^{\lambda}{}_{\mu\lambda\nu}  &  =-(n-1)Kg_{\mu\nu
},\nonumber\\
\bar{R}=g^{\mu\nu}\bar{R}{}_{\mu\nu}  &  =-n(n-1)K,\nonumber\\
\bar{R}^{\rho}{}_{\sigma\mu\nu}g^{\mu\tau}\bar{R}^{\sigma}{}_{\rho\tau
\upsilon}  &  =-2(n-1)K^{2}g_{\nu\upsilon},\nonumber\\
\bar{R}^{\rho}{}_{\sigma\mu\nu}g^{\mu\tau}g^{\nu\upsilon}\bar{R}^{\sigma}%
{}_{\rho\tau\upsilon}  &  =-2n(n-1)K^{2},\label{contractions}\\
R^{\rho}{}_{\sigma\mu\nu}g^{\mu\tau}\bar{R}^{\sigma}{}_{\rho\tau\upsilon}  &
=2KR_{\nu\upsilon},\nonumber\\
R^{\rho}{}_{\sigma\mu\nu}g^{\mu\tau}g^{\nu\upsilon}\bar{R}^{\sigma}{}%
_{\rho\tau\upsilon}  &  =2KR.\nonumber
\end{align}
Notice that some of these contractions are remarkable, especially the mixed
ones of the Riemann tensor with the MCT.

\section{A covariant derivative with both affine and Yang-Mills connections
and its lagrangian.}

In this Section we will construct a covariant derivative that allows us to
write derivatives in lagrangians that remain invariant under both
diffeomorphisms and Yang-Mills gauge transformations. We do this by the simple
expedient of defining a unified covariant derivative that is the sum of the
Yang-Mills and affine connections. Care should be taken with the fact that
there is no metricity, so that, even if it is still possible to use the metric
to lower or lift an index, it is \emph{false} that $B^{\nu}{}_{;\lambda
}=(B_{\mu}g^{\mu\nu})_{;\lambda}\overset{?}{=}B_{\mu;\lambda}g^{\mu\nu}$.

We begin by considering an affine connection $\Gamma_{\mu\nu}{}^{\lambda}$ and
with it defining the covariant derivative of a covariant vector%
\[
B_{\nu;\mu}=\partial_{\mu}B_{\nu}-\Gamma_{\mu\nu}{}^{\lambda}B_{\lambda}.
\]
It is possible to reexpress this covariant derivative in a different way using
an operator $\tilde{D}_{\mu},$ which we define by means of its components:%
\begin{equation}
(\tilde{D}_{\mu})_{\nu}{}^{\lambda}\equiv\partial_{\mu}\delta_{\nu}{}%
^{\lambda}-\Gamma_{\mu\nu}{}^{\lambda}. \label{LC derivative}%
\end{equation}
The interpretation of this formula is that the connection $\Gamma_{\mu}$ is a
matrix with indices $\nu$ and $\lambda.$ When this operator $\tilde{D}_{\mu}$
acts on a vector it results in the covariant derivative:%
\[
(\tilde{D}_{\mu})_{\nu}{}^{\lambda}B_{\lambda}=\partial_{\mu}\delta_{\nu}%
{}^{\lambda}B_{\lambda}-\Gamma_{\mu\nu}{}^{\lambda}B_{\lambda}=B_{\nu;\mu}.
\]

The similarity of (\ref{YM derivative}) and (\ref{LC derivative}) suggests a
unified covariant derivative that has both an affine and a Yang-Mills
connection:%
\begin{equation}
(\mathfrak{D}_{\mu})_{\nu}{}^{\lambda}\equiv\mathbf{1}_{N}\partial_{\mu}%
\delta_{\nu}{}^{\lambda}+\delta_{\nu}{}^{\lambda}A_{\mu}-\mathbf{1}_{N}%
\Gamma_{\mu\nu}{}^{\lambda}. \label{g derivative}%
\end{equation}
The symbol $\mathbf{1}_{N}$ was defined in the Conventions Section. The
unified covariant derivative $\mathfrak{D}_{\mu}$ is a matrix that has entries
both in the spacetime coordinates and in the internal space of a
representation of the compact Lie group of the Yang-Mills Theory. If it acts
on a vector $B_{\lambda}$ which is an element of the Lie algebra we get%
\begin{align}
(\mathfrak{D}_{\mu})_{\sigma}{}^{\lambda}B_{\lambda}  &  \equiv\mathbf{1}%
_{N}\partial_{\mu}\delta_{\sigma}{}^{\lambda}B_{\lambda}+\delta_{\sigma}%
{}^{\lambda}A_{\mu}B_{\lambda}-\mathbf{1}_{N}\Gamma_{\mu\sigma}{}^{\lambda
}B_{\lambda}\label{g vector}\\
&  =\partial_{\mu}B_{\sigma}+A_{\mu}B_{\sigma}-\Gamma_{\mu\sigma}{}^{\lambda
}B_{\lambda}.\nonumber
\end{align}
The symbols $\Gamma_{\mu\sigma}{}^{\lambda}$ and $\mathbf{1}_{N}$ commute
since the affine connection is not in the Lie algebra. On the other hand the
$A_{\mu}$ and the $B_{\sigma}$ do not commute, being both elements of that algebra.

This unified derivative transforms as%
\begin{equation}
\mathfrak{D}_{\mu}\rightarrow U\mathfrak{D}_{\mu}U^{-1} \label{g transform}%
\end{equation}
under gauge Yang-Mills transformations. This follows immediately from
(\ref{YM transform}) and the fact that the last term in $\mathfrak{D}_{\mu}$
commutes with the Yang-Mills transformation operators $U$. This way
$\mathfrak{D}_{\mu}$ transforms suggests postulating the following lagrangian
density:%
\begin{equation}
\mathcal{L}=-\frac{1}{2ng^{2}}g^{\mu\tau}g^{\nu\upsilon}\operatorname*{Tr}%
\,\overline{\operatorname*{Tr}}\mathfrak{D}_{[\mu}\cdot\mathfrak{D}_{\nu
]}\cdot\mathfrak{D}_{[\tau}\cdot\mathfrak{D}_{\upsilon]}, \label{lag1}%
\end{equation}
where besides the usual coefficient of Yang-Mills Theories $1/2g^{2}$ we have
included the reciprocal of the spacetime dimension $n,$ for reasons soon to be
clear. The $\operatorname*{Tr}$ is the usual Yang-Mills trace over the Lie
algebra and the $\overline{\operatorname*{Tr}}$ is a trace over the coordinate
indices of (\ref{g derivative}). (In $(\mathfrak{D}_{\mu})_{\sigma}{}%
^{\lambda}$ the $\sigma$ and $\lambda$ indicate the matrix indices.) This
lagrangian is invariant under diffeomorphisms because it is, by construction,
a scalar. It is also invariant under the Yang-Mills gauge transformation
(\ref{g transform}) since all the unitary operators $U$ and $U^{-1}$ cancel
among themselves.

Let us first evaluate the field strength curvature $\mathfrak{D}^{[\mu}%
\cdot\mathfrak{D}^{\nu]}.$ To do the calculation we must assume that the
partials are acting on some arbitrary function $\phi$ to their right, so that
we can replace $\partial_{\mu}A_{\nu}\phi-A_{\nu}\partial_{\mu}\phi
=(\partial_{\mu}A_{\nu})\phi.$ This way we obtain:%
\begin{align}
\lbrack(\mathfrak{D}_{\mu})_{\sigma}{}^{\upsilon},(\mathfrak{D}_{\nu
})_{\upsilon}{}^{\rho}]  &  =(\partial_{\mu}\delta_{\sigma}{}^{\upsilon
}+\delta_{\sigma}{}^{\upsilon}A_{\mu}-\Gamma_{\mu\sigma}{}^{\upsilon
})(\partial_{\nu}\delta_{\upsilon}{}^{\rho}+\delta_{\upsilon}{}^{\rho}A_{\nu
}-\Gamma_{\nu\upsilon}{}^{\rho})-\mu\leftrightarrow\nu\label{g curvature}\\
&  =\partial_{\lbrack\mu}A_{\nu]}\delta_{\sigma}{}^{\rho}+[A_{\mu},A_{\nu
}]\delta_{\sigma}{}^{\rho}+\Gamma_{\lbrack\mu\sigma,\nu]}{}^{\rho}%
+\Gamma_{\lbrack\mu\sigma}^{\upsilon}\Gamma_{\nu]\upsilon}^{\rho}\nonumber\\
&  =F_{\mu\nu}\delta_{\sigma}{}^{\rho}+\mathbf{1}_{N}R^{\rho}{}_{\sigma\mu\nu
}.\nonumber
\end{align}
That is, the commutator gives a sum of the Yang-Mills field strength tensor
and the Riemann tensor. To this we must add the curvature due to the metric
(\ref{Riemann2}). This term cannot come from any commutator of covariant
derivatives since it does not depend on any connection, just on the metric. We
must then add it and obtain the curvature obtained from connections:
\[
F_{\mu\nu}\delta_{\sigma}{}^{\rho}+\mathbf{1}_{N}R^{\rho}{}_{\sigma\mu\nu
}+\mathbf{1}_{N}\bar{R}^{\rho}{}_{\sigma\mu\nu}=F_{\mu\nu}\delta_{\sigma}%
{}^{\rho}+\mathbf{1}_{N}\hat{R}^{\rho}{}_{\sigma\mu\nu}.
\]
Here we have used definition (\ref{Riemann12}).

Let us go on and calculate in a more explicit form the lagrangian (\ref{lag1})
using the results just obtained. The $\overline{\operatorname{Tr}}$ is not
included because the initial and final indices of the right side of the
equation have been summed over:%
\begin{align}
\mathcal{L}  &  =\frac{1}{2ng^{2}}\operatorname*{Tr}\left(  (F_{\mu\nu}%
\delta_{\sigma}{}^{\rho}+\mathbf{1}_{N}\hat{R}^{\rho}{}_{\sigma\mu\nu}%
)g^{\mu\tau}g^{\nu\upsilon}(F_{\tau\upsilon}\delta_{\rho}{}^{\sigma
}+\mathbf{1}_{N}\hat{R}^{\sigma}{}_{\rho\tau\upsilon})\right) \nonumber\\
&  =\frac{1}{2g^{2}}\operatorname*{Tr}(F_{\mu\nu}^{2})+\frac{N}{2ng^{2}}%
\hat{R}^{\rho}{}_{\sigma\mu\nu}g^{\mu\tau}g^{\nu\upsilon}\hat{R}^{\sigma}%
{}_{\rho\tau\upsilon}\equiv\mathcal{L}_{B}+\mathcal{L}_{G} \label{lag2}%
\end{align}
Here $n$ is the dimensionality of spacetime and $N$ of the fundamental
representation of the Lie group. The mixed terms are zero since $R^{\sigma}%
{}_{\sigma\mu\nu}=0.$ The first term in the result, $\mathcal{L}_{B},$ is
simply the Yang-Mills gauge field lagrangian. The second term, $\mathcal{L}%
_{G}$, has to do with gravitation.

Let us get a quick approximate estimate of the coefficient $N/2ng^{2}$ using
frequently used values of the quantities involved, just to get an idea of the
size. Take as a grand unification group $SU(5),$ so $N=5$, and four dimensions
for spacetime, so $n=4.$ Assuming an asymptotic coupling constant
$\alpha(Q^{2}=\infty)=1/40$ and the usual relation $\alpha=g^{2}/4\pi,$ we get
$g^{2}=4\pi/40\approx3,$ so $N/2ng^{2}\approx1/5$. This coefficient has no
units and is roughly 1.

Let us assume a fermionic sector $\mathcal{L}_{F}$ in the theory of the form%
\begin{equation}
\mathcal{L}_{F}=\bar{\psi}\gamma^{\alpha}e_{\alpha}{}^{\mu}\mathbf{(1}%
_{N}\partial_{\mu}+\mathbf{1}_{N}\Gamma_{\mu}+A_{\mu})\psi\label{fermions}%
\end{equation}
where the fermion fields $\psi$ are chiral and transform in a complex
representation of the Lie group, and where the symbol $\Gamma_{\mu}$ is
defined in (\ref{spin con}). The correct fermionic lagrangian is a
symmetrization of the one above. I will not write the symmetrizations
explicitly in this paper to keep related equations and derivations simpler.
The correct \emph{symmetrized} result for the fermionic stress-energy tensor
is explicitly written in \cite{BD}. The total action of this theory is then
the sum of three terms: gravitational, bosonic and fermionic%
\begin{equation}
I_{T}=\int e(\mathcal{L}_{B}+\mathcal{L}_{G}+\mathcal{L}_{F})\mathcal{\,}%
d^{\,n}x \label{action}%
\end{equation}

Much of the point of this paper has to do with the solution to the equations
of motion generated by the first-order variations of the action $I$ with
respect to the metric $g_{\mu\nu}$ and the connection $\Gamma_{\mu\nu}%
{}^{\lambda}.$ There is a substantial amount of algebra involved in finding
these equations of motion. The algebraic development is given in the Appendix.
In next Section we take the equations from the Appendix and proceed directly
to solve them.

\section{First order solution of the two equations of motion: emergence of
General Relativity (GR).}

A perennial theoretical problem in quantum field theory is its clear
prediction of a huge vacuum energy due to quantum fluctuations. We know that
the virtual processes predicted by quantum field theory do exist, since their
contributions to experimentally observed quantities in the standard model of
high energy physics are necessary in order to have consistency with
experiment. On the other hand, for such a huge energy density, Einstein field
equation predicts a tiny universe. So we are in the peculiar situation of
having a clear-cut prediction from quantum field theory, the most successful
theory in physics, not being observed. Our aim here is to kill two birds with
one stone, by using this huge energy density as a large scale to be able to
perturbatively solve the equations of motion, while at the same time giving an
explanation for its inconspicuousness.

The vacuum energy acts as a constant energy density $\rho_{0}$, so that the
stress-energy tensor can be written in the form $T_{\rho\tau}=\rho_{0}%
g_{\rho\tau}+T_{\rho\tau}^{\prime}$, where $T_{\rho\tau}^{\prime}$ is the
stress-energy tensor with the vacuum energy subtracted. The stress-energy
tensor should satisfy $\left\langle 0|T_{\rho\tau}|0\right\rangle =\rho
_{0}g_{\rho\tau}.$ In the Appendix we have derived the two equations of motion
generated by action (\ref{action}) for the independent first variations of the
metric and the connection. Collecting these equations from the Appendix:%
\begin{align}
\frac{N}{ng^{2}}\left(  2S_{\rho\tau}-\frac{1}{2}Sg_{\rho\tau}+4K(R_{\rho\tau
}-\frac{1}{2}Rg_{\rho\tau})+K^{2}n(n-1)g_{\rho\tau}\right)   &  =\rho
_{0}g_{\rho\tau}+T_{\rho\tau}^{\prime},\label{metric}\\
\text{and\qquad}\frac{8N}{ng^{2}}\left(  eR^{\rho}{}_{\sigma\mu\nu}g^{\mu\tau
}g^{\nu\upsilon}+e\bar{R}^{\rho}{}_{\sigma\mu\nu}g^{\mu\tau}g^{\nu\upsilon
}\right)  _{;\upsilon}+e\bar{\psi}\gamma^{\tau}\mathbf{\sigma}^{\rho}%
{}_{\sigma}\psi &  =0, \label{connection}%
\end{align}
where we have expanded the stress-energy tensor into vacuum and real matter
parts. Notice the appearance of Einstein's equation left side in the first equation.

\subsection{Solution of the first equation of motion (\ref{metric}).}

It is accepted lore nowadays that, theoretically, the vacuum density should be
so large that%
\[
\left\langle T_{00}^{\prime}\right\rangle /|\rho_{0}|\sim10^{-123}.
\]
This estimate comes from assuming that $\left\langle T_{00}^{\prime
}\right\rangle $ is of the order of the critical density of the universe, and
that the vacuum energy density of the quantum fields would be given by some
quantum theory that uses Plank's constant, so that $\rho_{0}\sim G^{-2}.$ Let
us assume only that $\rho_{0}\gg|T_{00}^{\prime}|$ (but not that $\rho_{0}\sim
G^{-2})$ and use this relation to establish a large scale to solve the
equations of motion perturbatively.

In Eq. (\ref{metric}) there are two terms explicitly involving the metric, and
we assume that they cancel each, so that, up to order $K^{2}$,%
\begin{equation}
(n-1)\frac{N}{g^{2}}K^{2}=\rho_{0}. \label{density}%
\end{equation}
(Remember that $K$ is the constant in the MCT (\ref{Riemann2}).) Since $N\sim
n\sim g\sim1,$ we conclude that $K^{2}\sim\rho_{0},$ so that $K$ is huge, too,
with respect to the other curvature terms and $T_{\rho\tau}^{\prime}$. Having
established $K$ as a large scale, we get, to order $K,$ Einstein's equation%
\[
R_{\rho\tau}-\frac{1}{2}Rg_{\rho\tau}=-8\pi GT_{\rho\tau}^{\prime},
\]
assuming we have made the identification%
\begin{equation}
-8\pi G=\frac{ng^{2}}{4KN}. \label{gravitation}%
\end{equation}
(Here we are neglecting the terms $2S_{\rho\tau}-\frac{1}{2}Sg_{\rho\tau}$
because they consist of weak gravitational fields squared, and they are not
multiplied by the large factor $K,$ like the other terms of the left-hand side
of (\ref{metric}).) This identification makes sense since it implies
$K^{2}\sim G^{-2},$ thus verifying that $K$ is very large. Furthermore, these
two relations allow us the make the prediction%
\[
\rho_{0}=\frac{(n-1)n^{2}g^{2}}{1024\pi^{2}N}\cdot\frac{1}{G^{2}},
\]
or, order-of-magnitude, $\rho_{0}\sim G^{-2}.$ This relation is a prediction
of the model. We have obtained it using only the hypothesis that $\rho_{0}%
\gg|T_{00}^{\prime}|,$ no more.

\subsection{Solution to the second equation of motion (\ref{connection}).}

Let us pay attention now to the second equation of motion (\ref{connection}).
Let us write this equation in the form%
\begin{equation}
K\left(  eg_{\sigma\lbrack\mu}\delta_{\nu]}{}^{\rho}g^{\mu\tau}g^{\nu\upsilon
}\right)  _{;\upsilon}+\left(  eR^{\rho}{}_{\sigma\mu\nu}g^{\mu\tau}%
g^{\nu\upsilon}\right)  _{;\upsilon}+\frac{ng^{2}}{8N}e\bar{\psi}\gamma^{\tau
}\mathbf{\sigma}^{\rho}{}_{\sigma}\psi=0, \label{connection1}%
\end{equation}
where we have used (\ref{Riemann2}). In order for the equation to be
satisfied, considering that both the second and third terms are very small,
the coefficient of the very large scale factor $K$ has to be zero, or%
\begin{equation}
\left(  eg_{\sigma\mu}\delta_{\nu}{}^{\rho}g^{\mu\tau}g^{\nu\upsilon
}-eg_{\sigma\nu}\delta_{\mu}{}^{\rho}g^{\mu\tau}g^{\nu\upsilon}\right)
_{;\upsilon}=0. \label{m of m}%
\end{equation}
This equation implies the metricity (\ref{metricity}). A resulting equation is
left that relates the second and third terms of Eq. (\ref{connection1}) and
the spin of elementary particles with the divergence of Riemann's tensor. For
macroscopic purposes the spin term should be an expectation value,
$\left\langle \bar{\psi}\gamma^{\tau}\mathbf{\sigma}^{\rho}{}_{\sigma}%
\psi\right\rangle .$ The quantum vacuum expectation value of this term is
zero, since the spin contributions from all the particles should cancel. As
far as matter goes, the expectation value is simply a density average value.
For the solar system this term is basically negligible. However, for galaxies
with large quantities of interstellar gases and plasmas, especially in the
presence of galactic magnetic fields capable of polarizing them, there should
be a strong coupling of the polarized quantum spin with the divergence of the
Riemann curvature tensor, according to this model.

Perhaps it would be clarifying here to recall the situation when the Palatini
variation is taken in the case of GR. In this case (if one accepts the
presence of fermion fields and that the spin connection of these fields
depends on the affine connection), an equation similar to (\ref{connection1})
is found:%
\[
-\frac{1}{16\pi G}\left[  \frac{1}{2}(eg^{\sigma\nu})_{;\nu}\delta_{\tau
}^{\rho}+\frac{1}{2}(eg^{\rho\nu})_{;\nu}\delta_{\tau}^{\sigma}-(eg^{\sigma
\rho})_{;\tau}\right]  =e\bar{\psi}\gamma^{\rho}\mathbf{\sigma}^{\sigma}%
{}_{\tau}\psi.
\]
The large value of the coefficient $1/G$ on the left and the small value of
the quantity on the right assures us that the quantity in square brackets is
basically zero, a situation that can only occur if $g_{\mu\nu;\lambda}=0$.
Thus the manifold possesses metricity again, and we have obtained a theory
\emph{classically} similar to GR. Thus the Palatini variation with respect to
the connection in this model results in an equation similar to the same
variation in the standard metric-affine model.

Going back to Eqs. (\ref{connection1}) and (\ref{m of m}), in the absence of
matter one could simply set%
\begin{equation}
R^{\rho}{}_{\sigma\mu\nu}{}^{;\nu}=0, \label{div}%
\end{equation}
where we have used the metricity to simplify the equation. At first sight this
would seem to certainly result in a theory very different from GR. Actually,
this is not the case as we shall see. We will first derive an implication of
(\ref{div}) that is related to Einstein's equation. Take (\ref{div}) and
contract $\rho$ and $\mu$ to get $R_{\sigma\nu}{}^{;\nu}=0.$ Now the Bianchi
identities can be written in the form $R_{\sigma\nu}{}^{;\nu}-\frac{1}%
{2}g_{\sigma\nu}R^{,\nu}=0,$ so we conclude from them and the previous
equation that $R^{,\nu}=0$, that is, \emph{the curvature scalar has to be
constant}. This condition does not exist in GR in general but GR several
solutions upheld it anyway.

In the table that follows we list some representative solutions of Einstein's
equation, with their values for $R^{\lambda}{}_{\mu\nu\kappa;\lambda},$
$R^{,\nu},$ and $G{}_{\mu\nu}=R_{\mu\nu}-\frac{1}{2}g_{\mu\nu}R$ given for all
points except perhaps a null measure set:%
\[%
\begin{tabular}
[c]{|l|l|l|l|}\hline
\textbf{Solution} & $R^{\lambda}{}_{\mu\nu\kappa;\lambda}$ & $R^{,\nu}$ &
$G{}_{\mu\nu}$\\\hline\hline
Schwarzschild & 0 & 0 & 0\\\hline
Kerr & 0 & 0 & 0\\\hline
Reissner-Nordstrom & Not 0 & 0 & Not 0\\\hline
Kerr-Newman & Not 0 & 0 & Not 0\\\hline
Robertson-Walker inflation $\kappa=0$ & 0 & 0 & Not 0\\\hline
Robertson-Walker dust $\kappa=0$ & Not 0 & Not 0 & Not 0\\\hline
\end{tabular}
\]
Thus there is immediate agreement for three of those solutions of the GR with
the model presented here: Schwarzschild, Kerr, and Robertson Walker inflation
with flat space $\kappa=0$. In particular, there is agreement in the
Schwarzschild solution case, which accounts for most of the accurate
verifications of GR. There is no agreement for some of the solutions, but that
is not necessarily an argument against the model, since in those particular
cases we are not certain of the correction of the very large scale behavior of
GR. Those differences could be useful explaining a component of dark matter,
or finding a dynamical explanation for dark energy.

We thus arrive at a theory that is macroscopically similar (although not
identical) to GR, and one which enforces metricity.

\section{Quantization of the theory: its high energy limit.}

We have presented a model in this paper that has as a low energy limit GR. We
now study the problem of the renormalization of this theory. This is
essentially an ultraviolet problem so we will be concerned only with very high
energies $E\gg E_{P}.$ Yes, in this model Planck's energy is the low energy
limit. Our starting point is the original lagrangian (\ref{lag2}) of the
model, from where we get for pure gravity%
\begin{equation}
\mathcal{L}_{G}=\frac{N}{2ng^{2}}\left(  R^{\rho}{}_{\sigma\mu\nu}+\bar
{R}^{\rho}{}_{\sigma\mu\nu}\right)  g^{\mu\tau}g^{\nu\upsilon}\left(
R^{\sigma}{}_{\rho\tau\upsilon}+\bar{R}^{\sigma}{}_{\rho\tau\upsilon}\right)
. \label{lag3}%
\end{equation}
The scale of the energy of the $\bar{R}^{\rho}{}_{\sigma\mu\nu}$ term is
$|K|^{1/2}\sim G^{-1/2}\sim E_{P}.$ While this is certainly a large energy
scale we can consider even higher energies $E\gg E_{P}.$ In this very high
energy limit the independence of the metric and the connections is restored.
The point to notice here is that neither the MCT nor the Riemann tensor
contain derivatives of the metric. As a matter of fact there are not any
derivatives of the metric anywhere in the lagrangian. Thus for very high
energies the terms involving the connection keep gaining kinetic energy, while
terms involving the metric have stagnant energies. Eventually what is going to
occur is that $\left\vert R^{\rho}{}_{\sigma\mu\nu}\right\vert \gg\left\vert
\bar{R}^{\rho}{}_{\sigma\mu\nu}\right\vert .$ So we can simplify Eq.
(\ref{lag3}) and just write
\begin{equation}
\mathcal{L}_{G\text{ very high energy}}=\frac{N}{2ng^{2}}R^{\rho}{}_{\sigma
\mu\nu}g^{\mu\tau}g^{\nu\upsilon}R^{\sigma}{}_{\rho\tau\upsilon}. \label{lag4}%
\end{equation}
In this model the metric is not a quantum field, it is always classical. It
does not have a canonical conjugate field to form a quantum commutator with.
The quantum field of gravity in this model is the affine connection that makes
up the Riemann tensor. Its kinetic energy terms are%
\[
\mathcal{L}_{G}=\frac{N}{2ng^{2}}\left(  \Gamma_{\sigma\mu}{}^{\rho}{}_{,\nu
}-\Gamma_{\sigma\nu}{}^{\rho}{}_{,\mu}\right)  g^{\mu\tau}g^{\nu\upsilon
}\left(  \Gamma_{\rho\tau}{}^{\sigma}{}_{,\upsilon}-\Gamma_{\rho\upsilon}%
{}^{\sigma}{}_{,\tau}\right)
\]
Let us find the free equation of motion of the connection. We do this by
taking a first variation with respect to $\Gamma_{\sigma\mu}{}^{\rho}$:%
\[
\delta\int e\mathcal{L}_{G}d^{\,n}x=\frac{N}{ng^{2}}\int e\left(
\Gamma_{\sigma\mu}{}^{\rho}{}_{,\nu}-\Gamma_{\sigma\nu}{}^{\rho}{}_{,\mu
}\right)  g^{\mu\tau}g^{\nu\upsilon}\delta\Gamma_{\rho\tau}{}^{\sigma}%
{}_{,\upsilon}d^{\,n}x
\]
At these energies the metric is a classical background field changing at a far
smaller rate than the quantum excitations (the connections) and we simply take
it as constant. Applying parts we can then obtain the equation of motion the
connections obey:%
\begin{equation}
\Gamma_{\sigma\mu}{}^{\rho}{}_{,\nu}{}^{,\nu}-\Gamma_{\sigma\nu}{}^{\rho}%
{}_{,\mu}{}^{,\nu}=0. \label{eq motion}%
\end{equation}
We can take advantage of the diffeomorphism gauge invariance and set
$\Gamma_{\sigma\nu}{}^{\rho}{}_{,\mu}{}^{,\nu}=0,$ so that the equation simply
becomes the wave equation $\Gamma_{\sigma\mu}{}^{\rho}{}_{,\nu}{}^{,\nu}=0.$
The connections travel like other quantum fields.

For purposes of quantization the gauge invariance has to be dealt with in a
different fashion. One way of dealing with this problem is adding the
gauge-fixing term%
\[
\mathcal{L}_{\text{fix}}=-\frac{\lambda}{2}(\partial^{\mu}\Gamma_{\sigma\mu}%
{}^{\rho})(\partial^{\nu}\Gamma_{\rho\nu}{}^{\sigma})
\]
to the pure gravity lagrangian. Thus the matrix operator (in square brackets)
in the resulting lagrangian%
\[
\mathcal{L}_{G}+\mathcal{L}_{\text{fix}}=\frac{1}{2}\Gamma_{\sigma\mu}{}%
^{\rho}[g^{\mu\nu}\partial^{2}-\left(  1-\lambda\right)  \partial^{\mu
}\partial^{\nu}]\Gamma_{\rho\nu}{}^{\sigma}%
\]
is no longer singular (its singular nature is due to the diffeomorphism
invariance) and can be inverted, resulting in a convenient propagator. This,
plus the fact that the coupling constant has no units, gives very good
perspectives for quantization.

\section{Final comments.}

We have constructed a unified covariant derivative using a Yang-Mills
connection and an affine connection summed together. We then assume that the
affine connection and the metric are independent, and make the observation
that, just as the connection can generate curvature (measured by the Riemann
tensor), so could the metric produce a curvature (measured by the MCT, the
metric curvature tensor). A long but straightforward chain of mathematical
steps and the large scale of the vacuum energy due to quantum fields result,
in a low energy regime, in GR. For very high energies the MCT can be
disregarded and we get for pure gravity a theory very similar to a Yang-Mills
one, with the affine connections playing the role of vector gauge fields.
These connections are the quantum fields of gravity, and the metric is just a
background classical field at this energy level. The coupling constant has no
units and therefore the theory is probably renormalizable. Since it mimics the
Feynman diagram structure of a Yang-Mills Theory, it probably preserves
unitarity, too.

A comment about unitarity. When Levi-Civita connections enter in a lagrangian
a way similar to Yang-Mills fields, the metrics that make up these connections
appear in complicated products of powers of first and second order derivatives
of the metric,\cite{Stelle} greatly complicating the unitarity
issue.\cite{Phys Rep} In the model presented in this paper, at very high
energies all the metrics that appear are not being differentiated, and
therefore the metric is just a classical background field. During the
quantization process in the ultraviolet regime the uncomfortable powers of
derivatives of the metric simply do not appear at all.

Actually, the theory that appears in the low energy regime is not exactly GR.
As studied in Section 4 it has some solutions that are the same as GR, while
others are different. In particular, the Schwarzschild solution is the same
for both models. Since it is this solution that has most of the experimental
backing, this model does not have conflict with macroscopic observation. As a
matter of fact, its extra freedom in large structure solutions is welcome as
observation on large scales has resulted in a rather unclear experimental and
theoretical situations in GR, to witness, dark energy and dark
matter.\cite{review gravity}

This model is conceptually and mathematically simple, yet raises the
possibility of a quantizable gravity. It has the added merit of giving a
mechanism that allows for the vanishing of the vacuum energy density of the
quantum fields. It relates in a unequivocal logical way the large value of
this vacuum energy density and the weakness of $G,$ the gravitational
constant. It also gives a link between Yang-Mills Theories and gravity
(remember the Yang-Mills coupling constant $g$ is entering in gravitational
equations) but the relation seems incidental and does not shed light at all on
the origin of Yang-Mills Theories.

The arguments given in the previous Section on quantization explain why at
very high energies some terms dominate over others. If we assume very high
temperatures, like they existed during the Big Bang, it is possible to
introduce a chemical potential and the formalism of finite-temperature quantum
field theory. In this case the chemical potential increases with the
temperature and eventually the vacuum that is responsible for the system
acting as GR becomes metastable. When the system falls into the real vacuum of
the theory it will resemble the ultraviolet behavior studied in last
Section.\bigskip\bigskip

\noindent{\Large Appendix A. The derivation of the equations of motion from
the total lagrangian.\bigskip}

In this appendix we find the equations of motion that result from taking a
first variation with respect to the affine connection and the metric of the
total action of this model (\ref{action}):%
\begin{equation}
I_{T}=I_{G}+I_{M} \label{total action}%
\end{equation}
where%
\[
I_{G}=\frac{N}{2ng^{2}}\int e\hat{R}^{\rho}{}_{\sigma\mu\nu}g^{\mu\tau}%
g^{\nu\upsilon}\hat{R}^{\sigma}{}_{\rho\tau\upsilon}d^{\,n}x=\frac{N}{2ng^{2}%
}\int e\hat{S}d^{\,n}x
\]
and%
\begin{equation}
I_{M}=\int e(\mathcal{L}_{B}+\mathcal{L}_{F})\mathcal{\,}d^{\,n}x.
\label{action matter}%
\end{equation}
Recall the definition of $\hat{R}^{\rho}{}_{\sigma\mu\nu}$ is given by
(\ref{Riemann12}) and of $\hat{S}$ by (\ref{defs}).

\emph{1) We consider first the variation of }$I_{T}$\emph{ with respect to the
metric.}

We proceed by calculating the first variation of $e\hat{S}$ with respect to
the metric.\textit{ }The metric appears in $e;$ as usual $\delta e=-\frac
{1}{2}eg_{\mu\tau}\delta g^{\mu\tau},$ so that the variation in $e$ due to the
metric is%
\begin{align*}
\hat{S}\delta e  &  =-\frac{1}{2}e\hat{S}g_{\mu\tau}\delta g^{\mu\tau}\\
&  =-\frac{1}{2}e\left(  S+4RK-2K^{2}n(n-1)\right)  g_{\mu\tau}\delta
g^{\mu\tau},
\end{align*}
where we have used the identities in (\ref{contractions}). The metric also
appears in the symbol $\hat{S}$ of Eq. (\ref{defs}), explicitly performing the
product between the two curvature tensors, and the variation in $\hat{S}$ due
to this presence is%
\begin{align*}
\delta\hat{S}_{1}  &  =2\hat{S}_{\mu\tau}\delta g^{\mu\tau}\\
&  =2\left(  S_{\mu\tau}+4KR_{\mu\tau}-2K^{2}(n-1)g_{\mu\tau}\right)  \delta
g^{\mu\tau},
\end{align*}
where we have used again identities (\ref{contractions}). The last place the
metric appears is in the MCT $\bar{R}^{\rho}{}_{\sigma\mu\nu}$ (but it never
appears in the $R^{\rho}{}_{\sigma\mu\nu}$). This first variation can be
calculated as follows. Use $\delta g_{\mu\nu}=-g_{\mu\rho}g_{\nu\sigma}\delta
g^{\rho\sigma}$ to verify that%
\[
\delta\bar{R}^{\rho}{}_{\sigma\mu\nu}=K\delta_{\nu}{}^{\rho}\delta
g_{\sigma\mu}-K\delta_{\mu}{}^{\rho}\delta g_{\sigma\nu}=-K\delta_{\lbrack\nu
}{}^{\rho}g_{\mu]\lambda}g_{\sigma\upsilon}\delta g^{\upsilon\lambda}.
\]
With this result prove that%
\begin{align}
\delta\hat{S}_{2}  &  =2\delta\bar{R}^{\rho}{}_{\sigma\mu\nu}g^{\mu\tau}%
g^{\nu\upsilon}(R^{\sigma}{}_{\rho\tau\upsilon}+\bar{R}^{\sigma}{}_{\rho
\tau\upsilon})\nonumber\\
&  =-4K\left(  R{}_{\mu\tau}-(n-1)Kg{}_{\mu\tau}\right)  \delta g^{\mu\tau}.
\label{3-term}%
\end{align}
The total variation is given $\delta(e\hat{S})=\hat{S}\delta e+e\delta\hat
{S}_{1}+e\delta\hat{S}_{2}.$ To find the equation of motion we must consider
the variation of $I_{M}$ with respect to the metric. This can be done using
(\ref{stress}) and (\ref{stress-f}), and setting%
\[
T_{\mu\nu}=-\frac{2}{e}\frac{\delta I_{M}}{\delta g^{\mu\nu}},
\]
where $I_{M}$ is given by (\ref{action matter}). The sum of the gravitational
and the matter variations gives:%
\[
\frac{N}{2ng^{2}}\int\delta(e\hat{S})d^{\,n}x-\frac{1}{2}\int eT_{\mu\nu
}\delta g^{\mu\nu}=0,
\]
from which we obtain the equation of motion generated by the metric:%
\begin{equation}
\frac{N}{ng^{2}}\left(  2S_{\rho\tau}-\frac{1}{2}Sg_{\rho\tau}+4K(R_{\rho\tau
}-\frac{1}{2}Rg_{\rho\tau})+K^{2}n(n-1)g_{\rho\tau}\right)  =T_{\rho\tau}.
\label{eq-1}%
\end{equation}

\emph{2) We consider now the first variation of }$I_{T}$\emph{ with respect to
the connection. }The connection appears in $I_{T}$ only in two places: in the
Riemann curvature tensor, and in the spin connection that appears in $I_{M}$
in the lagrangian $\mathcal{L}_{F}$ shown in (\ref{fermions}). The spin
connection (\ref{con}) contains the affine connection in the covariant
derivative. As before, we proceed first to study the variation of $e\hat{S},$
in this case with respect to the connection.

Notice that%
\[
\delta\hat{R}^{\sigma}{}_{\rho\tau\upsilon}=\delta R^{\sigma}{}_{\rho
\tau\upsilon}+\delta\bar{R}^{\sigma}{}_{\rho\tau\upsilon}=\delta R^{\sigma}%
{}_{\rho\tau\upsilon}=(\delta\Gamma_{\rho\lbrack\tau}{}^{\sigma})_{;\upsilon
]}.
\]
The first variation of the MCT is zero, $\delta\bar{R}^{\sigma}{}_{\rho
\tau\upsilon}=0,$ since it is a functional of the metric only, and the last
step in the Eq. above relies on Palatini's identity. Then the variation of the
gravitational term of the actions is:%
\begin{align}
\delta I_{G}  &  =\frac{N}{ng^{2}}\int e\hat{R}^{\rho}{}_{\sigma\mu\nu}%
g^{\mu\tau}g^{\nu\upsilon}(\delta\Gamma_{\rho\lbrack\tau}{}^{\sigma
})_{;\upsilon]}d^{\,n}x,\nonumber\\
&  =-2\frac{N}{ng^{2}}\int\left(  e\hat{R}^{\rho}{}_{\sigma\mu\nu}g^{\mu\tau
}g^{\nu\upsilon}\right)  _{;\upsilon}\delta\Gamma_{\rho\tau}{}^{\sigma}%
d^{\,n}x, \label{parts}%
\end{align}
where in the last step we have used \textquotedblleft covariant
parts\textquotedblright.

We now confront the variation of the matter term $I_{M}$ of the action. (The
lagragian density $\mathcal{L}_{B}$ is assumed to consist of Yang-Mills and
similar theories that do not depend on the affine connection. In Yang-Mills
Theories all the covariant derivatives appear in the form $A_{[\mu;\nu
]}=A_{[\mu,\nu]},$ that does not contain connections.) We calculate this
variation through the functional derivative%
\begin{align*}
\frac{\delta I_{M}}{\delta\Gamma_{\rho\tau}{}^{\sigma}}  &  =\int\frac{\delta
e\mathcal{L}_{F}}{\delta\Gamma_{\rho\tau}{}^{\sigma}}d^{\,n}x=\int e\bar{\psi
}\gamma^{\mu}\frac{\delta\Gamma_{\mu}}{\delta\Gamma_{\rho\tau}{}^{\sigma}}\psi
d^{\,n}x\\
&  =\frac{1}{4}\int e\bar{\psi}\gamma^{\mu}\sigma^{\alpha}{}_{\beta}%
\frac{\delta e_{\alpha}{}^{\lambda}e^{\beta}{}_{\lambda;\mu}}{\delta
\Gamma_{\rho\tau}{}^{\sigma}}\psi d^{\,n}x=-\frac{1}{4}e\bar{\psi}\gamma
^{\tau}\mathbf{\sigma}^{\rho}{}_{\sigma}\psi,
\end{align*}
using the definitions (\ref{con}) and (\ref{spin con}) and the convention of
letters from the beginning and the middle of the Greek alphabet set up in
Section 2. With this result and our previous (\ref{parts}) we get the equation
of motion generated by the connection:%
\begin{equation}
\frac{8N}{ng^{2}}\left(  e\hat{R}^{\rho}{}_{\sigma\mu\nu}g^{\mu\tau}%
g^{\nu\upsilon}\right)  _{;\upsilon}=-e\bar{\psi}\gamma^{\tau}\mathbf{\sigma
}^{\rho}{}_{\sigma}\psi. \label{eq-2}%
\end{equation}

We have obtained the two coupled equations of motion that result from a first
variation of the action (\ref{total action}) with respect to the connection
and the metric. In Section 4 above we find an interesting perturbative
solution to them.


\begin{thebibliography}{9}                                                                                                %


\bibitem {Deser}S. Deser 2000 Infinities in quantum gravities \emph{Annalen
Phys.} \textbf{9} 299-307
\href{http://arxiv.org/abs/gr-qc/9911073}{\emph{Preprint }arXiv:gr-qc/9911073}

\bibitem {correction}N. E. J. Bjerrum-Bohr et al. 2003 Quantum corrections to
the Schwarzschild and Kerr metrics \emph{Phys. Rev. }\textbf{D68} 084005-084035

\bibitem {EFT}C. P. Burgess 2007 Introduction to effective field theory, to be
published in \emph{Annual Reviews of Nuclear and Particle Science }1-55
\href{http://arxiv.org/abs/hep-th/0701053v2}{\emph{Preprint }%
arXiv:hep-th/0701053v2}

\bibitem {Querella}L. Querella 1998 Variational principles and cosmological
models in higher-order gravity \emph{Doctoral dissertation IAGL Universit\'{e}
de Li\`{e}ge} 1-183 \href{http://arxiv.org/abs/gr-qc/9902044}{\emph{Preprint
}arXiv:gr-qc/9902044}

\bibitem {BD}N. D. Birrell and P. C. W. Davies 1982 \emph{Quantum fields in
curved space time }(Cambridge: Cambridge University Press)

\bibitem {Stelle}K. S. Stelle 1977 Renormalization of higher-derivative
quantum gravity \emph{Phys. Rev.} \textbf{D16} 953-969

\bibitem {Phys Rep}F. W. Hehl et al. 1995 Metric-affine gauge theory of
gravity: field equations, Noether identities, world spinors, and breaking of
dilation invariance \emph{Phys. Rep.} \textbf{258} 1-171

\bibitem {review gravity}M. Farhoudi 2006 On higher order gravities, their
analogy to GR, and dimensional dependent version of Duff's trace anomaly
relation \emph{Gen. Rel. Grav.} \textbf{38} 1261-1284
\href{http://arxiv.org/abs/physics/0509210}{\emph{Preprint }%
arXiv:physics/0509210}
\end{thebibliography}
\end{document}